\documentclass[aps,prd,showpacs,superscriptaddress,ctexart,nofootinbib,onecolumn]{revtex4}

\usepackage{amsmath}
\usepackage{graphicx}
\usepackage{dcolumn}
\usepackage{bm}
\usepackage{amssymb}
\usepackage{latexsym}
\usepackage{simplewick}
\usepackage{slashed,graphicx,color,amsmath,amssymb,simplewick}
\usepackage{axodraw}

\usepackage{color}

\begin{document}

\title{IR Divergence in Inflationary Tensor Perturbations from Fermion Loops}

\author{Kaixi Feng\footnote{Email: fengkaixi10@mails.gucas.ac.cn}}
\affiliation{College of Physical Sciences, Graduate School of Chinese Academy of Sciences, Beijing 100049, China}

\author{Yi-Fu Cai\footnote{Email: ycai21@asu.edu}}
\affiliation{Department of Physics, Arizona State University, Tempe, AZ 85287, USA}
\affiliation{Department of Physics, McGill University, Montr\'eal, QC, H3A 2T8, Canada}

\author{Yun-Song Piao\footnote{Email: yspiao@gucas.ac.cn}}
\affiliation{College of Physical Sciences, Graduate School of Chinese Academy of Sciences, Beijing 100049, China}

\begin{abstract}
We estimate fermion loop corrections to the two-point correlation function of primordial tensor perturbations in a slow-roll inflationary background. We particularly compute an explicit term of one-loop correction from a massless fermion, and then extend to the complete Interaction Hamiltonian. After that, we study one-loop corrections contributed by a massive fermion to primordial tensor fluctuations. The loop correction arisen from a massless fermion field contains logarithms and thus may constrain the validity of perturbation theory in inflationary cosmology, but the situation could be relaxed once the fermion's mass is taken into account. Another one-loop diagram for a massive fermion which involves one vertex is constrained by a UV cutoff as expected by quantum field theory. Our result shows that loop corrections of a fermion field have the same sign as those of a scalar field, and thus implies that the inclusion of fermion loop corrections may not help to alleviate the issue of IR divergence in inflationary cosmology.
\end{abstract}


\maketitle


\section{Introduction}

Inflation is considered as the most successful model of describing physics of very early universe, which has explained conceptual issues of Big Bang cosmology \cite{Guth:1980zm, Linde:1981mu, Albrecht:1982wi} (see \cite{Starobinsky:1980te, Fang:1980wi, Sato:1980yn} for early works). Among these remarkable achievements, inflation has predicted a nearly scale-invariant primordial power spectrum which was later verified in high precision by Cosmic Microwave Background (CMB) observations \cite{Komatsu:2010fb}. The success of inflation is mainly based on a series of assumptions including an enough long period of quasi-exponential expansion and the applicability of perturbation theory during this phase. Therefore, an important question which ought to be understood is whether there exists a bound beyond which the perturbation theory would break down in inflationary cosmology.

Generally, the primordial perturbations during inflation were originated from quantum fluctuations with their physical wavelengths being much shorter than the Hubble radius for which the modes are in ultraviolet (UV) regime. Along with the quasi-exponential expansion, the wavelengths of these modes would be stretched outside the Hubble radius for which the modes are in infrared (IR) regime. Through this mechanism quantum fluctuations are able to become classical ones and finally can be observed in today's CMB experiments (see \cite{Mukhanov:1990me} for an overview of cosmological perturbation theory). Therefore, the evolution of metric perturbation can be calculated at the moment of Hubble crossing until the end of inflation. This mechanism applies to both the scalar type and tensor type of metric perturbation during inflation. Especially, for tensor fluctuations the amplitude of one degree of freedom is determined in order of $H/M_p$, and this result is not sensitive to the details of inflation models.

Since the observable modes of inflationary fluctuations are in IR regime, the reliability of perturbation theory expects that these modes may behave well and two-point correlation functions may be finite even when higher order corrections are included. However, as was noticed by a pioneer work \cite{Sasaki:1992ux} long time ago, perturbation theory beyond leading order had troublesome IR behavior in a de Sitter space. Another issue of IR effect which may affect late time evolution of the universe due to the accumulation of long-wavelength inflationary fluctuations was studied in \cite{Tsamis:1992sx, Tsamis:1993ub, Abramo:1997hu, Mukhanov:1996ak}.

In the past ten years, following Maldacena's calculation of three-point functions in a slow-roll inflation model \cite{Maldacena:2002vr}, there have been significant theoretical developments on refining our understanding of quantum field description of inflationary fluctuations. A remarkable work systematically studying quantum effect to arbitrary order in cosmological perturbations was recently formulated by Weinberg \cite{Weinberg:2005vy} (see \cite{Kahya:2010xh} for further studies of the time dependence of the correlator.). With an example of massless scalar loop corrections, it was observed that the momentum dependence of two-point correlation function of inflationary fluctuations evolves as a logarithmic function in IR limit \cite{Weinberg:2005vy}. This result implies that perturbation theory may break down since of the contribution of IR fluctuations to loop integrals \cite{Riotto:2008mv, Burgess:2009bs, Xue:2011hm, Garbrecht:2011gu}. There have been a lot of studies on the effects of IR divergences on scalar type curvature perturbations in inflation (see \cite{Seery:2010kh} for a recent review; and see \cite{Weinberg:2006ac, Sloth:2006az, Chaicherdsakul:2006ui, Sloth:2006nu, Seery:2007we, Seery:2007wf, Bartolo:2007ti, Urakawa:2008rb, Enqvist:2008kt, Dimastrogiovanni:2008af, Adshead:2008gk, Campo:2009fx, Kumar:2009ge, Senatore:2009cf, Byrnes:2010yc, Xue:2012wi, Chen:2012ye} for detailed studies of various inflationary loop corrections; and also see \cite{Prokopec:2002uw, Brunier:2004sb, Boyanovsky:2005sh, Tsamis:2005je, Miao:2005am, Kahya:2007bc} for fully dimensionally regulated and renormalized computations in inflation.). Note that, the IR effect of primordial inflationary perturbations in the presence of only adiabatic modes could be very limit \cite{Urakawa:2010it, Gerstenlauer:2011ti, Riotto:2011sf, Giddings:2011zd}. Therefore, the study of IR issue is much more interesting when entropy modes are taken into account, namely, a second scalar or a fermionic field minimally coupled to inflationary background.

In this paper, we estimate the IR issue arising from loop corrections of a fermionic field to the two-point correlation function of primordial tensor fluctuations in slow-roll inflationary cosmology. We first consider loop contribution from a massless fermion. As an exercise, we calculate one term of the Interaction Hamiltonian and find that it contributes to one loop involving two vertices and gives rise to a logarithmic correction to the two-point correlation function of tensor fluctuations. After extending to the complete form of the Interaction Hamiltonian, this logarithmic correction remains while only the coefficient in front of it is improved. However, once we consider the fermion's mass is not zero, we find the logarithmic function would be replace by a $\cosh$ function and thus the IR issue could be much relaxed. In addition, for a massive fermion there exists another one-loop diagram which only involves one vertex, but this diagram only contributes to the UV regime and thus is not of enough observable interest. Our main result shows that the loop corrections contributed by a fermion field have the same sign as those contributed by a scalar field, and thus it implies that the inclusion of fermion loop corrections cannot be used to alleviate the issue of IR divergence in inflationary cosmology.

The present paper is organized as follows. Section II gives a brief introduction to the background dynamics of a fermion field in inflationary cosmology. In Section III, we first simply review the canonical quantization of linear fluctuation modes in cosmological background, and then introduce the {\it in-in} formalism as a setup for computing loop diagrams. After that, we expand the inflationary Lagrangian involving a fermionic field to the third and forth order and thus derive the interaction Hamiltonian for a fermionic field coupled to tensor fluctuations straightforwardly. Making use of the {\it in-in} formalism and the interaction Hamiltonian, we in Section IV estimate loop corrections of both a massless and a massive fermionic field to the two-point correlation function of primordial tensor fluctuations up to one-loop order. Finally, we conclude with a discussion in Section V. Note that, we use the natural units $8\pi G=\hbar=c=1$ and all parameters are normalized by $M_p=1/\sqrt{8 \pi G}$ in this paper.

\section{Background Dynamics}

To begin with, we simply review the background dynamics of a spinor field which is minimally coupled to Einstein's gravity (see Refs. \cite{Weinberg, BirrellDavies} for detailed introduction and see \cite{ArmendarizPicon:2003qk, Cai:2008gk} for recent phenomenological study in cosmology). Following the general covariance principle, a connection between the metric $g_{\mu\nu}$ and the vierbein is given by
\begin{equation}
 g_{\mu\nu}e_{a}^{\mu}e_{b}^{\nu}=\eta_{ab}~,
\end{equation}
where $e_{a}^{\mu}$ denotes the vierbein, $g_{\mu\nu}$ is the space-time metric, and $\eta_{a b}$ is the Minkowski metric with $\eta_{ab}={\rm diag}(1,-1,-1,-1)$. Note that the Latin indices represents the local inertial frame and the Greek indices represents the space-time frame.

We choose the Dirac-Pauli representation as
\begin{eqnarray}
 \gamma^0= \left(\begin{array}{cccc}
 1 &   0  \\
 0 &   -1
 \end{array}\right),~~~
 \gamma^{i}=\left(\begin{array}{cccc}
 0 &          \sigma_{i} \\
 -\sigma_{i}&   0
 \end{array}\right),
\end{eqnarray}
where $\sigma_{i}$ is Pauli matrices. One can see that the $4\times4$ $\gamma^{a}$ satisfy the Clifford algebra $\{\gamma^{a},\gamma^{b}\}=2\eta_{ab}$. The $\gamma^{a}$ and $e_{a}^{\mu}$ provide the definition of a new set of Gamma matrices
\begin{equation}
 \Gamma^{\mu}=e_{a}^{\mu}\gamma^{a}~,
\end{equation}
which satisfy the algebra $\{\Gamma^{\mu},\Gamma^{\nu}\}=2g^{\mu\nu}$. The generators of the Spinor representation of the Lorentz group can be written as $\Sigma^{ab}=\frac{1}{4}[\gamma^{a},\gamma^{b}]$. So the covariant derivative are given by
\begin{eqnarray}
 D_{\mu}\psi&=&(\partial_{\mu}+\Omega_{\mu})\psi~,\\
 D_{\mu}\bar\psi&=&\partial_{\mu}\bar\psi-\bar\psi\Omega_{\mu}~,
\end{eqnarray}
where the Dirac adjoint is defined as $\bar\psi\equiv\psi^+\gamma^0$. The $4\times4$ matrix $\Omega_{\mu} = \frac{1}{2}\omega_{\mu ab}\Sigma^{ab}$ is the spin connection, where $\omega_{\mu ab} = e_{a}^{\nu} \nabla_{\mu}e_{\nu b}$ are Ricci spin coefficients.

By the aid of the above algebra we can write down the following Dirac action in a curved space-time background
\begin{eqnarray}\label{action}
 S_{\psi} = \int d^4 x~e~ [\frac{i}{2}(\bar\psi\Gamma^{\mu}D_{\mu} \psi-D_{\mu}\bar\psi\Gamma^{\mu}\psi)-m\bar\psi\psi]~.
\end{eqnarray}
Here, $e$ is the determinant of the vierbein $e_{\mu}^{a}$ and $m$ stands for the mass of the spinor field $\psi$.

Varying the action with respect to the vierbein $e_{a}^{\mu}$, we obtain the energy-momentum-tensor,
\begin{eqnarray}\label{EMT}
 T_{\mu\nu}&=&\frac{e_{\mu a}}{e}\frac{\delta S_\psi}{\delta e_{a}^{\nu}} \nonumber\\
 &=& \frac{i}{4} [\bar\psi\Gamma_{\nu}D_{\mu}\psi +\bar\psi \Gamma_{\mu}D_{\nu}\psi -D_{\mu}\bar\psi\Gamma_{\nu}\psi -D_{\nu}\bar\psi\Gamma_{\mu}\psi] -g_{\mu\nu}{\cal L}_{\psi}~.
\end{eqnarray}
On the other hand, varying the action with respect to the field $\bar\psi$, $\psi$ respectively yields the following equations of motion,
\begin{eqnarray}
 i\Gamma^{\mu}D_{\mu}\psi-m\psi &=& 0~,\\
 iD_{\mu}\bar\psi\Gamma^{\mu}+m\bar\psi &=& 0~.
\end{eqnarray}

We deal with the homogeneous and isotropic FRW metric,
\begin{equation}
 ds^{2}=dt^{2}-a^{2}(t)d\vec{x}^2~.
\end{equation}
Correspondingly, the vierbein are given by
\begin{equation}
 e_{0}^{\mu}=\delta_{0}^{\mu}~,~~e_{i}^{\mu}=\frac{1}{a}\delta_{i}^{\mu}~.
\end{equation}
Assuming the spinor field is space-independent, the equation of motion reads $ i\gamma^{0} (\dot{\psi} + \frac{3}{2}H\psi) - m \psi = 0 $, where a dot denotes a derivative with respect to the cosmic time and $H$ is the Hubble parameter. Taking a further derivative, we can obtain:
\begin{equation}\label{solution}
 \bar\psi\psi=\frac{N}{a^{3}}~,
\end{equation}
where $N$ is a positive time-independent constant.

From the result (\ref{solution}), we can find the background dynamics of a massive spinor field behaves like the cold dark matter with its energy density goes as $a^{-3}$. If the universe is in inflationary phase, the energy density of a spinor field will be diluted immediately and cannot contribute to the background evolution. Therefore, in the quantum treatment of cosmological perturbations in later context, we can safely choose $\langle\psi\rangle=0$ as the vacuum state for the fermion field during inflation.

\section{Setup of perturbation analysis}

Inflationary cosmology usually assumed a homogeneous scalar field with a non-zero expectation value varying slowly along its potential. This background dynamics is characterized by a series of slow-roll parameter, among which the most significant parameter is given by $\epsilon\equiv-\frac{\dot{H}}{H^2}$. In this scenario, we naturally have cosmological perturbations of both scalar and tensor types. The interaction between the scalar perturbation and a fermion was studied comprehensively in \cite{Chaicherdsakul:2006ui}. In the current paper, we particularly focus on the interaction between tensor fluctuation and a fermion field.

\subsection{Canonical quantization of linear fluctuations}

Taking into account the spatial dependence of the fermion field, we can write down the equations of motion for the fermion and tensor fluctuations as follows,
\begin{eqnarray}
\label{EoMpsi}
 i\gamma^0(\dot\psi+\frac{3}{2}H\psi)-i\gamma^i\frac{\partial_i}{a}\psi -m\psi &=&0~, \\
\label{EoMhij}
 \ddot h_{ij} +3H\dot h_{ij} -\frac{\partial_i^2}{a^2}h_{ij} &=&0~.
\end{eqnarray}

Moreover, we make the Fourier transformations of these fields, which are given by
\begin{eqnarray}
 \psi(\mathbf{x},t) = \int\mathrm{d}^3p \sum_s \left[ e^{i\mathbf{p}\cdot\mathbf{x}} \alpha_{\mathbf{p},s} X_{\mathbf{p},s}(t) +e^{-i\mathbf{p}\cdot\mathbf{x}} \beta^\dag_{\mathbf{p},s} W_{\mathbf{p},s}(t) \right]~,
\end{eqnarray}
and
\begin{eqnarray}
 h_{ij}(\mathbf{x},t) = \int\mathrm{d}^3q \sum_\lambda \left[ e^{i\mathbf{q}\cdot\mathbf{x}} \epsilon_{ij}(\hat{q},\lambda) \alpha(\mathbf{q},\lambda) h_q(t) +e^{-i\mathbf{q}\cdot\mathbf{x}} \epsilon_{ij}^\ast(\hat{q},\lambda) \alpha^\ast(\mathbf{q},\lambda) h_q^\ast(t) \right]~,
\end{eqnarray}
where $s=\pm\frac{1}{2}$ is the spin of the fermion, $\lambda=\pm2$ stands for the helicity of tensor fluctuation, and $\epsilon_{ij}(\hat{q},\lambda)$ is the polarization of tensor mode with the following normalization \cite{Weinberg:2008zzc}:
\begin{eqnarray} \label{eq:porlarization}
 \sum_{\lambda=\pm2} \epsilon_{ij}(\hat{q},\lambda) \epsilon^\ast_{kl}(\hat{q},\lambda) &=& \delta_{ik}\delta_{jl} +\delta_{il}\delta_{jk} -\delta_{ij}\delta_{kl} +\delta_{ij}\hat{q}_k\hat{q}_l +\delta_{kl}\hat{q}_i\hat{q}_j \nonumber\\
 && -\delta_{ik}\hat{q}_j\hat{q}_l -\delta_{il}\hat{q}_j\hat{q}_k -\delta_{jk}\hat{q}_i\hat{q}_l -\delta_{jl}\hat{q}_i\hat{q}_k +\hat{q}_i\hat{q}_j\hat{q}_k\hat{q}_l~,
\end{eqnarray}
where $\hat{q}$ is the unit vector along the $\mathbf{q}$ direction.

The annihilation operators $\alpha_{\mathbf{p},s}$, $\beta_{\mathbf{p},s}$, $\alpha(\vec{q},\lambda)$ satisfy the following commutation relations
\begin{eqnarray}
 \{\alpha(\mathbf{p},s), \alpha^\dag(\mathbf{p'},s')\} = \{\beta(\mathbf{p},s), \beta^\dag(\mathbf{p'},s')\} = \delta_{ss'} \delta^3(\mathbf{p}-\mathbf{p}') ~,
\end{eqnarray}
\begin{eqnarray}
 \{\alpha(\mathbf{p},s), \alpha(\mathbf{p'},s')\} = \{\beta(\mathbf{p},s), \beta(\mathbf{p'},s')\} = 0 ~,
\end{eqnarray}
and
\begin{eqnarray}
 \left[\alpha(\mathbf{q}, \lambda), \alpha^\ast(\mathbf{q}', \lambda')\right] &=& \delta_{\lambda\lambda'} \delta^3(\mathbf{q}-\mathbf{q}') ~,\\
 \left[\alpha(\mathbf{q}, \lambda), \alpha(\mathbf{q}', \lambda')\right] &=& 0 ~,
\end{eqnarray}
respectively.

Additionally, the Fourier modes $X_{\mathbf{p},s}(t)$, $W_{\mathbf{p},s}(t)$ and $h_q(t)$ satisfy the equations of motion for the fermion and tensor fluctuation in cosmological background as shown in Eqs. (\ref{EoMpsi}) and (\ref{EoMhij}) respectively. At quadratic order, each Fourier mode evolves independently. Therefore, we can solve the equations of motion for those fluctuations in comoving time coordinate $\tau\equiv\int dt/a$, and obtain
\begin{eqnarray}
\label{sol_psi}
 X_{k,\pm}(\tau) &=& \frac{i\sqrt{-\pi k\tau}}{2(2\pi a)^{3/2}} e^{\pm\frac{\pi m}{2H}} H^{(1)}_{\frac{1}{2}\mp i\frac{m}{H}}(-k\tau) ~,\\
\label{sol_h}
 h_k(\tau) &=& \frac{\sqrt{16\pi G}}{(2\pi)^{3/2}}\frac{H}{\sqrt{k^3}}(1+ik\tau)e^{-ik\tau}~,
\end{eqnarray}
for the fermion and tensor mode, respectively. Note that, in above solutions we have made use of Bunch-Davies vacuum as initial condition to determine the coefficient of the Hankel function, and the convention of primordial tensor modes is the same as that adopted in \cite{Cai:2007xr}. Also we notice that, when the fermion is massless (m=0), the solutions to the fermion modes are independent of the helicity and take the form of $\frac{e^{-ik\tau}}{\sqrt{2}(2\pi a)^{3/2}}$, which is asymptotically the plane wave function.

\subsection{Interactions}

After obtaining solutions of linear fluctuations, we then study the interaction terms for loop calculation. Since the Hamiltonian that governs primordial fluctuations is time dependent in our case, we need to make use of {\it in-in} formalism. Following Weinberg's formula \cite{Weinberg:2005vy}, we can calculate the {\it in-in} correlation function to arbitrary order via the following expression,
\begin{eqnarray}\label{eq:in-in-2}
 \langle Q(t)\rangle=\sum_{N=0}^\infty i^N\int_{-\infty}^t\mathrm{d}t_N\int_{-\infty}^{t_N}\mathrm{d}t_{N-1}...\int_{-\infty}^{t_2}\mathrm{d}t_1
 \langle[H_I(t_1),[H_I(t_2),...[H_I(t_N),Q^I(t)]...]]\rangle ~,
\end{eqnarray}
where $H_I$ is the interaction Hamiltonian as will be discussed later. In this formula, the LHS is working in the interactive vacuum and the expectation value on the RHS is calculated in the effectively free field vacuum by annihilation operators working on cosmological background. Moreover, the index ``N" in Eq. \eqref{eq:in-in-2} connects the {\it in-in} formalism with the number of vertices in a Feynman diagram. For example, $N=0$ always represents propagators at tree level; $N=1$ refers to diagrams with only one vertex; and $N=2$ means that there are two vertices involved in each of these diagrams; and so on.

As is well known, in the {\it in-in} formalism the Hamiltonian is decomposed into a effectively free part $H_0$ which is of quadratic order in fluctuation variables and an interaction part $H_I$. First, we study the interaction Hamiltonian in cubic order. By expanding the action up to next-to-leading order, we get
\begin{eqnarray}\label{eq:int-action}
 \mathcal{S}_{int}=\frac{1}{2}\int\mathrm{d}^4x~\sqrt{-g}~T^{\mu\nu}~\delta g_{\mu\nu}
\end{eqnarray}
Since the scalar and tensor fluctuations are decoupled in tree level, we can write down the perturbed FRW metric only involving the tensor modes as follows,
\begin{eqnarray}
 \mathrm{d}s^2=a^2(\tau)(\mathrm{d}\tau^2-g_{ij}\mathrm{d}x^i\mathrm{d}x^j)
\end{eqnarray}
where
\begin{eqnarray}
 g_{ij}=\delta_{ij}+h_{ij}~,
\end{eqnarray}
with the tensor perturbation $h_{ij}$ satisfying the relations $\partial_i h_{ij}=0$ and $h_{ii}=0$. To first order in $h_{ij}$, Eq. \eqref{eq:int-action} becomes
\begin{eqnarray}\label{eq:int-action-massless}
 \mathcal{S}_{int}^3 = \frac{i}{8}\int\mathrm{d}^4x ~a^2 \left[ \bar{\psi}\gamma^i(\partial_j\psi) +\bar{\psi}\gamma^j(\partial_i\psi) -(\partial_i\bar{\psi})\gamma^j\psi -(\partial_j\bar{\psi})\gamma^i\psi \right] h_{ij}~.
\end{eqnarray}
Similarly, we can expand the action to the forth order and obtain one extra interaction term in order of $O(h^2)$. This is a coupling between the mass term of the fermion and two tensor modes, which is given by
\begin{eqnarray}\label{eq:int-action-massive}
 \mathcal{S}_{int}^4 = \frac{1}{4} \int\mathrm{d}^4x ~a^3 m\bar{\psi}\psi h_{ij}h_{ji}~.
\end{eqnarray}
We will discuss loop corrections arisen from these interaction terms one by one in later context.

\section{Fermion loop corrections to correlation function of tensor fluctuations}

In the previous section, we have derived out two interaction terms (\ref{eq:int-action-massless}) and (\ref{eq:int-action-massive}) which correspond to the cases of two fermions coupling to one tensor mode and two tensor modes, respectively. Thus, one can immediately read that there exist two different Feynman diagrams with one-loop correction which are depicted in the following figures.
\begin{center} \begin{figure}
\begin{picture}(300,100)(0,0)
\CArc(150,50)(40,0,180)
\CArc(150,50)(40,180,360)
\Photon(50,50)(110,50){5}{4} \Vertex(110,50){2}
\Photon(190,50)(250,50){5}{4} \Vertex(190,50){2}
\end{picture}\\
\caption{The two-point correlation function of primordial tensor fluctuations with one-loop fermion correction involving two vertices.}\label{Fig:one-loop-two-vertices}
\end{figure}
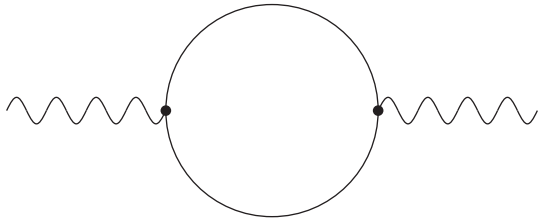
\end{center}
\begin{center}
\begin{figure}
\begin{picture}(300,100)(0,0)
\CArc(150,41.2132)(30,315,225)
\Vertex(150,3){2}\Photon(100,0)(200,0){4}{8.7}
\Curve{(128.787,20.2132)(150,3)}\Curve{(171.213,20.2132)(150,3)}
\end{picture}\\
\caption{The two-point correlation function of primordial tensor fluctuations with one-loop fermion correction only involving single vertex.}\label{Fig:one-loop-one-vertex}
\end{figure}
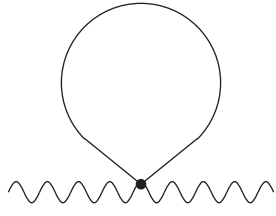
\end{center}
%

%
%
%

\subsection{A Massless Fermion}

As an exercise, we first consider that the fermion is massless. This is because at early times of our universe, the energy scale is extremely high and the masses of most particles are much lighter than that of inflaton. In this case, the fermion loop corrections can only be contributed by Fig. \ref{Fig:one-loop-two-vertices} since the interaction appeared in Eq. (\ref{eq:int-action-massive}) automatically vanishes when $m=0$.

\subsubsection{A illustrate of one-loop correction}

We first focus on one interaction term appeared in Eq. (\ref{eq:int-action-massless}), which leads to the following Interaction Hamiltonian,
\begin{eqnarray}\label{eq:part-of-hamiltionian}
 H_I = \frac{i}{8} \int\mathrm{d}^3x ~a^2 (\partial_i\bar{\psi}) \gamma^j\psi h_{ij}~.
\end{eqnarray}
According to the formula Eq. \eqref{eq:in-in-2}, the correlation function with one-loop fermion correction as depicted in Fig. \ref{Fig:one-loop-two-vertices} can be written as
\begin{eqnarray}\label{eq:sub-in}
 &~& \langle h_{mn}(\mathbf{x},t) h_{mn}(\mathbf{x}',t) \rangle_2 \nonumber\\
 &=& -\int_{-\infty}^t\mathrm{d}t_2 \int_{-\infty}^{t_2}\mathrm{d}t_1 \langle [H_I(t_1), [H_I(t_2), h_{mn}(t)h_{mn}(t) ]] \rangle \nonumber\\
 &=& \frac{1}{32} \int\mathrm{d}^3x_1\mathrm{d}^3x_2 \int_{-\infty}^t\mathrm{d}t_2 \int_{-\infty}^{t_2}\mathrm{d}t_1 p'^kp^i~ {\rm Re} \bigg[ \langle \bar{\psi}(t_1)\gamma^l\psi(t_1) \bar{\psi}(t_2)\gamma^j\psi(t_2) \rangle \nonumber\\
 &~& \hspace{1cm} \times ~ ( \langle h_{kl}(t_1)h_{ij}(t_2)h_{mn}(t)h_{mn}(t) \rangle - \langle h_{kl}(t_1)h_{mn}(t)h_{mn}(t)h_{ij}(t_2) \rangle ) \bigg]~.
\end{eqnarray}
Note that, in the above formula there is no extra minus sign from the partial operator in Eq. \eqref{eq:part-of-hamiltionian}, since the contractions always take place between the modes $e^{i\mathbf{p}\cdot\mathbf{x}}$ and $e^{-i\mathbf{p}\cdot\mathbf{x}}$. After contracting the particle modes according to Eq. \eqref{eq:sub-in}, then we get the following three useful expressions:
\begin{eqnarray}\label{eq:sub-1}
 &~&\langle h_{kl}(t_1)h_{ij}(t_2)h_{mn}(t)h_{mn}(t)\rangle \nonumber\\
 &=&2\int\mathrm{d}^3q_1\mathrm{d}^3q_1' \sum_\lambda e^{i\mathbf{q}_1\cdot(\mathbf{x}_1-\mathbf{x})+i\mathbf{q}_1'\cdot(\mathbf{x}_2-\mathbf{x}')} \epsilon_{kl}(\hat{q}_1,\lambda) \epsilon_{mn}^\ast(\hat{q}_1,\lambda) \epsilon_{ij}(\hat{q}_1',\lambda') \epsilon_{mn}^\ast(\hat{q}_1',\lambda') \nonumber\\
 &~& \hspace{3.4cm} \times ~ h_{q_1}(t_1) h_{q_1}^\ast(t) h_{q_1'}(t_2) h_{q_1'}^\ast(t) ~,
\end{eqnarray}
\begin{eqnarray}
\label{eq:sub-2}
 &~& \langle h_{kl}(t_1)h_{mn}(t)h_{mn}(t)h_{ij}(t_2)\rangle \nonumber\\
 &=& 2\int\mathrm{d}^3q_1\mathrm{d}^3q_1'\sum_\lambda e^{i\mathbf{q}_1\cdot(\mathbf{x}_1-\mathbf{x})+i\mathbf{q}_1'\cdot(\mathbf{x}_2-\mathbf{x}')} \epsilon_{kl}(\hat{q}_1,\lambda) \epsilon_{mn}^\ast(\hat{q}_1,\lambda) \epsilon_{mn}(-\hat{q}_1',\lambda') \epsilon_{ij}^\ast(-\hat{q}_1',\lambda') \nonumber\\
 &~& \hspace{3.4cm} \times ~ h_{q_1}(t_1) h_{q_1}^\ast(t) h_{q_1'}(t) h_{q_1'}^\ast(t_2) ~,
\end{eqnarray}
\begin{eqnarray}
\label{eq:sub-3}
 &~& \langle \psi^\ast(t_1)\psi(t_1)\psi^\ast(t_2)\psi(t_2)\rangle \nonumber\\
 &=& \int\mathrm{d}^3p\mathrm{d}^3p' e^{i(\mathbf{p}+\mathbf{p}')\cdot(\mathbf{x}_1-\mathbf{x}_2)}
 \sum_{s,s'} X_{\mathbf{p},s}(t_1) X^\ast_{\mathbf{p},s}(t_2) W_{\mathbf{p}',s'}(t_2) W^\ast_{\mathbf{p}',s'}(t_1) ~.
\end{eqnarray}

To substitute Eqs. (\ref{eq:sub-1}, \ref{eq:sub-2}, \ref{eq:sub-3}) into Eq. \eqref{eq:sub-in}, we finally have
\begin{eqnarray}\label{eq:sub-in-1}
 &~& \int\mathrm{d}^3x e^{i\mathbf{q}\cdot(\mathbf{x}-\mathbf{x}')} \langle h_{mn}(\mathbf{x},t) h_{mn}(\mathbf{x}',t) \rangle_2 \nonumber\\
 &=& \frac{(2\pi)^3}{64} \int_{-\infty}^\tau\mathrm{d}\tau_2 \int_{-\infty}^{\tau_2}\mathrm{d}\tau_1 {\rm Re} \int\mathrm{d}^3p\mathrm{d}^3p' \delta^3(\mathbf{p}+\mathbf{p}'+\mathbf{q}) \frac{p'^kp^i \mbox{tr} (\gamma^lp\!\!\!\slash\gamma^jp'\!\!\!\!\slash) }{pp'} \nonumber\\
 &~& \times ~\sum_{\lambda,\lambda'}\epsilon_{kl}(\hat{q},\lambda) \epsilon_{mn}^\ast(\hat{q},\lambda) \epsilon_{mn}(\hat{q},\lambda') \epsilon_{ij}^\ast(\hat{q},\lambda') e^{-i(p+p')\tau_1+i(p+p')\tau_2} \nonumber\\
 &&  \times ~h_{q}(t_1) h_{q}^\ast(t) ~[ h_{q}(t_2)h_{q}^\ast(t) -h_{q}(t)h_{q}^\ast(t_2) ] ~,
\end{eqnarray}
where we have made use of conformal time $\tau$ instead of the cosmic time $t$ via $ \tau \equiv \int^t \frac{\mathrm{d}t}{a} $. In addition, we have also applied a useful relation of the polarization tensor which is given by,
\begin{eqnarray}
 \epsilon^\ast_{ij}(\hat{q},\lambda)=\epsilon_{ij}(-\hat{q},\lambda)~.
\end{eqnarray}

Notice that, one can formulate the Fourier modes of a massless fermion as follows,
\begin{eqnarray}
 X_{\mathbf{p},s}(t)=\frac{1}{a^{3/2}(t)}u_{\mathbf{p},s}~, ~~~ W_{\mathbf{p,s}}(t)=\frac{1}{a^{3/2}(t)}v_{\mathbf{p},s}~,
\end{eqnarray}
where
\begin{eqnarray}
 u_{\mathbf{p},s}=u_{p,s}^0 e^{-ip\tau},\quad v_{\mathbf{p},s}=v_{p,s}^0 e^{ip\tau},
\end{eqnarray}
with $u^0$ and $v^0$ in the Minkowski space normalized as
\begin{eqnarray}
 \sum_s u_{p,s}^0\bar{u}_{p,s}^0 = \sum_s v_{p,s}^0\bar{v}_{p,s}^0 = \frac{\gamma^\mu p_\mu}{2(2\pi)^3p}
\end{eqnarray}
Since the fluctuations in the interacting picture are viewed as free fields, $h_q(t)$ is positive-frequency mode of which the form is provided in Eq. \eqref{sol_h}. Integrating over the conformal time in Eq. \eqref{eq:sub-in-1} yields the final result:
\begin{eqnarray}\label{eq:sub-in-2}
 &~& \int\mathrm{d}^3x e^{i\mathbf{q}\cdot(\mathbf{x}-\mathbf{x}')} \langle h_{mn}(\mathbf{x},t) h_{mn}(\mathbf{x}',t) \rangle_2 \nonumber\\
 &=& \frac{1}{16(2\pi)^3M_p^4} \int\mathrm{d}^3p\mathrm{d}^3p' \delta^3(\mathbf{p}+\mathbf{p}'+\mathbf{q}) \frac{p'^kp^i \mbox{tr}(\gamma^lp\!\!\!\slash\gamma^jp'\!\!\!\!\slash) }{pp'}\nonumber\\
 &~& \times ~ \sum_{\lambda,\lambda'} \epsilon_{kl}(\hat{q},\lambda) \epsilon_{mn}^\ast(\hat{q},\lambda) \epsilon_{mn}(\hat{q},\lambda') \epsilon_{ij}^\ast(\hat{q},\lambda') \frac{H^4}{q^6} \frac{-1}{2q(p+p'+q)}~.
\end{eqnarray}
With the useful normalization Eq. \eqref{eq:porlarization}, one can easily do the contraction between momentum and tensor polarization.

\subsubsection{The Complete Hamiltonian}

After having studied the one-loop correction from a single interaction term between the massless fermion and gravitational waves, then we extend the interaction Hamiltonian into the complete form:
\begin{eqnarray}\label{eq:hamiltonian}
 H_I = -\frac{i}{8} \int\mathrm{d}^3x ~a^2 [\bar{\psi}\gamma^i(\partial_j\psi) +\bar{\psi}\gamma^j(\partial_i\psi) -(\partial_i\bar{\psi})\gamma^j\psi -(\partial_j\bar{\psi})\gamma^i\psi] h_{ij}~,
\end{eqnarray}
which is only contributed by the perturbed action at cubic order. After a process of lengthy computation, we obtain the two-point correlation function for tensor modes as follows,
\begin{eqnarray} \label{eq:sub-4}
 &~& \int\mathrm{d}^3x e^{i\mathbf{q}\cdot(\mathbf{x}-\mathbf{x}')} \langle h_{mn}(\mathbf{x},t)h_{mn}(\mathbf{x}',t) \rangle_2 \nonumber\\
 &=& \frac{1}{4(2\pi)^3M_p^4} \int\mathrm{d}^3p \mathrm{d}^3p' \delta^3(\mathbf{p}+\mathbf{p}'+\mathbf{q})
    \frac{\mbox{tr}(\gamma^lp\!\!\!\slash\gamma^jp'\!\!\!\!\slash)(p^k-p'^k)(p^i-p'^i)}{pp'}\nonumber\\
 &~& \times ~ \sum_{\lambda,\lambda'} \epsilon_{kl}(\hat{q},\lambda) \epsilon_{mn}^\ast(\hat{q},\lambda) \epsilon_{mn}(\hat{q},\lambda') \epsilon_{ij}^\ast(\hat{q},\lambda') \frac{H^4}{q^6}\frac{1}{2q(p+p'+q)} ~,
\end{eqnarray}
where $\mathbf{q}$ is the momentum of the external line in Fig. \ref{Fig:one-loop-two-vertices}, and $\mathbf{p}$, $\mathbf{p}'$ are momenta corresponding to the two internal lines in the same figure. After contracting all the Lorentz indices, Eq. \eqref{eq:sub-4} can be further simplified as
\begin{eqnarray} \label{eq:loop-2}
 \int\mathrm{d}^3x e^{i\mathbf{q}\cdot(\mathbf{x}-\mathbf{x}')} \langle h_{mn}(\mathbf{x},t) h_{mn}(\mathbf{x}',t) \rangle_2
 = -\frac{H^4(t_q)}{2 (2\pi)^3 M_p^4 q^7} \times \frac{2\pi}{q}\mathcal{K}(q)~,
\end{eqnarray}
where
\begin{eqnarray}
 \mathcal{K}(q) \equiv \int_0^\infty\mathrm{d}p \int_{|p-q|}^{p+q}\mathrm{d}p'
 ~ \frac{(p'^2-p^2-q^2) (p^2-p'^2-q^2) }{ q^4(p+p'+q)}
 ~ \bigg[ 4p^2q^2-(p'^2-p^2-q^2)^2 \bigg] ~,
\end{eqnarray}
and $t_q$ denotes the time scale when the tensor mode with comoving wave number $q$ crosses the Hubble radius.

By virtue of the dimensional regularization analysis, the remaining momentum integral will lead to the following correspondence:
\begin{eqnarray}\label{eq:F-delta}
 \frac{2\pi}{q} \mathcal{K}(q) \to q^{4+\delta} F(\delta) ~,
\end{eqnarray}
where $\delta$ denotes the dimensional difference. When $\delta\to 0$, we have
\begin{eqnarray}\label{eq:F_0}
 \mathcal{K}(q) = \frac{1}{2\pi}q^5 (F_0\ln q + L) ~,
\end{eqnarray}
with $L$ being a divergent constant. In order to calculate the coefficient $F_0$, we differentiate $\mathcal{K}(q)$ and get,
\begin{eqnarray}
 F_0=\frac{608\pi}{15} ~.
\end{eqnarray}
In addition, the dimensional difference $\delta$ brings a log correction to the solutions of fermion and tensor modes, $X_{p,s}$ and $h_{q}$ (see \cite{Senatore:2009cf} for detailed study). To take into account this effect, and to combine the analysis of dimensional regularization, we eventually obtain the two-point correlation function of primordial tensor fluctuations under one-loop massless fermion corrections, which is given by,
\begin{eqnarray}\label{eq:massless-loop-final}
 \int\mathrm{d}^3x e^{i\mathbf{q}\cdot(\mathbf{x}-\mathbf{x}')} \langle h_{mn}(\mathbf{x},t) h_{mn}(\mathbf{x}',t) \rangle_2
 \rightarrow -\frac{304\pi H^4(t_q)}{15 (2\pi)^3 M_p^4 q^3} \times \ln\left(\frac{H(t_q)}{\mu}\right)~,
\end{eqnarray}
where $\mu$ is a physical renormalization scale. Note that, the form of our result is similar to the result of Ref. \cite{Chaicherdsakul:2006ui} in which the author studied the loop correction of a fermion field to the primordial curvature perturbation. Our result shows that the amplitude of the fermionic loop correction to primordial tensor fluctuation is much larger than that to the scalar type curvature perturbation. In addition, we notice that our result is obviously invariant under the following rescaling: $a\to\lambda a$, $x\to x/\lambda$, and $k\to\lambda k$. This conclusion is consistent with the analysis made in \cite{Senatore:2009cf}.

From our result, we notice that the sign of the loop correction arisen from a massless fermion is the same as that of a massless scalar field. This conclusion is in agreement with the result of Ref. \cite{Chaicherdsakul:2006ui}, in which the loop correction of a fermion field to curvature perturbation was analyzed. Here we would like to comment a little bit on this issue. It is well known that, there is always a negative sign contributed by a fermionic loop correction in quantum field theory. In quantum field theory the loop contribution is calculated based on the {\it in-out} formalism and the Wick's theorem guarantees that for a closed fermion loop, and thus we have the following relation,
\begin{eqnarray}\label{eq:qft-contraction}
 \underbrace{\bar{\psi}_1 \overbrace{\psi_1\bar{\psi}_2} \overbrace{\psi_2\bar{\psi}_3} ... \overbrace{\psi_{n-1}\bar{\psi}_n}\psi_n}
 = -\mbox{tr} \left[ \overbrace{\psi_1\bar{\psi}_2} \overbrace{\psi_2\bar{\psi}_3} ... \overbrace{\psi_{n-1}\bar{\psi}_n} \overbrace{\bar{\psi}_1\psi_n} \right] ~,
\end{eqnarray}
which explains the origin of a negative sign of the fermion loop. However, the perturbation theory in the cosmological background is based on the {\it in-in} formalism, and correspondingly the contraction relation is replaced by
\begin{eqnarray}\label{eq:in-in-contraction}
 \underbrace{\bar{\psi}_1 \overbrace{\psi_1\bar{\psi}_2} \overbrace{\psi_2\bar{\psi}_3} ... \overbrace{\psi_{n-1}\bar{\psi}_n} \psi_n}
 = \overbrace{\psi_1\bar{\psi}_2} \overbrace{\psi_2\bar{\psi}_3} ... \overbrace{\psi_{n-1}\bar{\psi}_n} \overbrace{\psi_n\bar{\psi}_1} ~,
\end{eqnarray}
since there is no time-order in this formalism.

\subsection{A Massive fermion}

After having analyzed the specific case of massless fermion loop corrections to the correlation function of primordial tensor fluctuations, we continue to study the loop effects brought by a massive fermion in this subsection. In the case of massive fermion, there exists another interaction due to the expansion of the mass term in the action. This interaction could bring a new loop correction with single vertex as shown in Fig. \ref{Fig:one-loop-one-vertex}.

\subsubsection{One-loop correction with two vertex}

We first consider the same interaction Hamiltonian as what has been studied in the case of a massless fermion. Its Feynmann diagram is shown in \eqref{Fig:one-loop-two-vertices}. We can simply repeat the process of previous calculation, but need to replace the plane wave-function of the massless fermion by the Hankel function which is the mode function for the massive fermion field.

In this case, the correlation function of primordial tensor fluctuations with one-loop fermion correction can be expressed as,
\begin{eqnarray}\label{massive1}
 &~& \int\mathrm{d}^3x e^{i\mathbf{q}\cdot(\mathbf{x}-\mathbf{x}')}
 \langle h_{mn}(\mathbf{x},t)h_{mn}(\mathbf{x}',t) \rangle_2 \nonumber\\
 &=& \frac{1}{16} \int\mathrm{d}^3x e^{i\mathbf{q}\cdot(\mathbf{x}-\mathbf{x}')} \int\mathrm{d}^3x_1 \mathrm{d}^3x_2 \int_{-\infty}^t\mathrm{d}t_2 a^2(t_2) \int_{-\infty}^{t_2} \mathrm{d}t_1 a^2(t_1)p'^kp^i~
    {\rm Re} \int\mathrm{d}^3p \mathrm{d}^3p' e^{i(\mathbf{p}+\mathbf{p}')\cdot(\mathbf{x}_1-\mathbf{x}_2)} \nonumber\\
 &~& \times ~ \sum_{s,s'} \gamma^l X_{\mathbf{p},s}(t_1) \bar{X}_{\mathbf{p},s}(t_2) \gamma^j W_{\mathbf{p}',s'}(t_2) \bar{W}_{\mathbf{p}',s'}(t_1) \int\mathrm{d}^3q_1 \mathrm{d}^3q'_1 \sum_\lambda e^{i\mathbf{q}_1\cdot(\mathbf{x}_1-\mathbf{x})+i\mathbf{q}_1'\cdot(\mathbf{x}_2-\mathbf{x}')} \nonumber\\
 &~& \times ~ \epsilon_{kl}(\hat{q}_1,\lambda) \epsilon_{mn}^\ast(\hat{q}_1,\lambda) \epsilon_{mn}(-\hat{q}_1',\lambda') \epsilon_{ij}^\ast(-\hat{q}_1',\lambda') h_{q_1}(t_1) h_{q_1}^\ast(t) [h_{q'_1}(t_2)h_{q'_1}^\ast(t)-h_{q'_1}(t)h_{q'_1}^\ast(t_2)]~, \nonumber\\
\end{eqnarray}
where the mode function $X_{\mathbf{p},s}$ is given by \eqref{sol_psi}. Making use of a little mathematical features of Hankel functions and the Dirac algebra, one can get the following useful relation,
\begin{eqnarray}\label{massive2}
 &~& \sum_{s,s'} \gamma^lX_{\mathbf{p},s}(t_1) \bar{X}_{\mathbf{p},s}(t_2) \gamma^jW_{\mathbf{p}',s'}(t_2) \bar{W}_{\mathbf{p}',s'}(t_1) \nonumber\\
 &=& \frac{2|\Gamma(\mu)|^4}{(2\pi)^8a^{3}(t_1)a^{3}(t_2)} \bigg(\cosh[\frac{2\pi m}{H}]-1\bigg) \bigg[ 2 -(\hat{p}\cdot\hat{p}')(p^{\frac{2im}{H}}p'^{-\frac{2im}{H}} + c.c. )  \bigg] \delta^{jl} + O(\varepsilon^{jl})
\end{eqnarray}
where we have introduced the index $\mu \equiv \frac{1}{2}-i\frac{m}{H}$ with $m$ being the mass of the fermion. The second term $O(\varepsilon^{jl})$ in rhs of \eqref{massive2} is asymmetric and thus will vanish due to symmetry. In addition, we have assumed the perturbation modes are super-Hubble scale where the Hankel function is approximately a power law function. Therefore, the following calculation is based on the assumption that all the modes of interests are super-Hubble.

Then we insert the relation \eqref{massive2} into \eqref{massive1}. After a process of lengthy calculation, we further obtain the simplified form of the two-point correlator as follows,
\begin{eqnarray}\label{massive3}
 &~& \int\mathrm{d}^3x e^{i\mathbf{q}\cdot(\mathbf{x}-\mathbf{x}')} \langle h_{mn}(\mathbf{x},t) h_{mn}(\mathbf{x}',t) \rangle_{2} \nonumber\\
 &=& -\frac{4 H^4|\Gamma(\mu)|^4}{(2\pi)^4M_p^4q^{11}} \bigg(\cosh[\frac{2\pi m}{H}]-1\bigg) \int_0^q\mathrm{d}p \int_{|p-q|}^{p+q}\mathrm{d}p'
 \left[ 4p^2q^2 -(p'^2-p^2-q^2)^2 \right] \nonumber\\
 &~&
 \times~ \bigg[ 2pp' -(\mathbf{p}\cdot\mathbf{p}')(p^{\frac{2im}{H}}p'^{-\frac{2im}{H}} + c.c. )  \bigg] ~,
\end{eqnarray}
where we have contracted the indices of tensor modes in detailed calculation. In addition, we can make use of the useful relation: $|\Gamma(\mu)|^2 = {\pi}/{\cosh[\frac{\pi m}{H}]}$. Then we do the integral of \eqref{massive3} very carefully, and find the final result is composed of some regular functions and many annoying hypergeometric functions.

In order to grasp the physics of this integral, we would like to express the final result in approximate forms under two different limits. First, in the limit of small $q$ value and small fermion mass $m$, the result of the integral is given by
\begin{eqnarray}\label{massive_a}
 &~& \int\mathrm{d}^3x e^{i\mathbf{q}\cdot(\mathbf{x}-\mathbf{x}')} \langle h_{mn}(\mathbf{x},t) h_{mn}(\mathbf{x}',t) \rangle_{2}|_{q\ll{aH},m\ll H}
 \simeq -\frac{536 H^2 m^2}{315M_p^4\cosh^2[\frac{\pi m}{H}]q^3} ~,
\end{eqnarray}
up to the leading order. Second, we still consider $q$ to be small but a much larger value of $m$, and then we get the asymptotic form as follow,
\begin{eqnarray}\label{massive_b}
 \int\mathrm{d}^3x e^{i\mathbf{q}\cdot(\mathbf{x}-\mathbf{x}')} \langle h_{mn}(\mathbf{x},t) h_{mn}(\mathbf{x}',t) \rangle_{2}|_{q\ll{aH},m \gg H} 
 \simeq -\frac{8H^4}{15\pi^2M_p^4 q^3} \frac{\sinh[\frac{2\pi m}{H}]\tan[\frac{\pi m}{H}]}{\cosh^2[\frac{\pi m}{H}]} ~.
\end{eqnarray}

From the above result, we can get the following information. For the case of a massive fermion, its one-loop correction to the correlation function of primordial tensor fluctuations does not change the momentum dependence and thus the primordial power spectrum is always scale-invariant. Moreover, when the fermion mass is very small, the contribution of the fermion loop is suppressed by the factor $m^2/H^2$; while the fermion mass is very large, the contribution of the fermion loop is suppressed exponentially due to the factor $1/\cosh^2[\frac{\pi m}{H}]$. Therefore, one may conclude that the logarithmic divergence of the fermion loop appeared in the case of massless fermion might be removed by the mass term. However, we should keep in mind that this result is based on the assumption that all primordial perturbation modes of interest are on super-Hubble scale and thus we have neglected the contribution of the loop integrals inside the Hubble scale during inflation. It is interesting to go to details on this issue and dig out the relation between the logarithmic divergence and the loop integrals related to sub-Hubble modes. We would like to leave it to future study.

\subsubsection{One-loop correction with single vertex}

Then we turn our attention to the calculation of one-loop correction of a massive fermion field with a single vertex. The relevant Feynmann diagram is shown in Fig. \ref{Fig:one-loop-one-vertex}, and its interaction term is given by \eqref{eq:int-action-massive}.

According to the formula developed in Ref. \cite{Weinberg:2005vy}, as there is only one vertex in Fig. \eqref{Fig:one-loop-one-vertex}, it corresponds to the $N=1$ term to Eq. \eqref{eq:in-in-2} with one communicator in the vacuum expectation. Thus, we have the following expression
\begin{eqnarray}\label{eq:massive-loop-1}
 &~& \langle h_{mn}(\mathbf{x},t)h_{mn}(\mathbf{x}',t) \rangle_{1}\nonumber\\
 &=& i\int_{-\infty}^t\mathrm{d}t_1 \langle [H_I(t_1),h_{mn}(t)h_{mn}(t)] \rangle \nonumber\\
 &=& \frac{i}{4}\int\mathrm{d}^3x_1\mathrm{d}^3x_2\int_{-\infty}^t\mathrm{d}t_1~a^3(t_1)~m~ \langle\bar{\psi}(t_1)\psi(t_1)\rangle \nonumber\\
 &~& \times ~ \bigg[ \langle h_{mn}(t)h_{mn}(t)h_{ij}(t_1)h_{ji}(t_1) \rangle - \langle h_{ij}(t_1)h_{ji}(t_1)h_{mn}(t)h_{mn}(t) \rangle \bigg]~.
\end{eqnarray}

Recall that we have $\langle\bar{\psi}(t_1)\psi(t_1)\rangle = \int\mathrm{d}^3p \sum_s \bar{W}_{\mathbf{p},s}(t_1) W_{\mathbf{p},s}(t_1)$ for the fermion field. Again, we make use of the Dirac algebra and contract the indices of tensor modes. After a little lengthy computation, we eventually get the result of the above integral:
\begin{eqnarray}\label{eq:massive-loop-final}
 &~& \int\mathrm{d}^3x e^{i\mathbf{q}\cdot(\mathbf{x}-\mathbf{x}')} <h_{mn}(\mathbf{x},t)h_{mn}(\mathbf{x}',t)>_{1} \nonumber\\
 &\simeq& \frac{64\pi mH^4}{(2\pi)^4M_p^4q^6} {\rm Im}
 \bigg[ \int_{-\infty}^{\tau} \mathrm{d}\tau_1 a(\tau_1) e^{-2iq\tau_1+2iq\tau}
    \int^{a(\tau_1)\Lambda_{UV}} \mathrm{d}^3p ~ {\bm 1} \bigg] \nonumber\\
 &\rightarrow& \frac{64m\Lambda_{UV}^3}{9\pi M_p^4 Hq^3}\bigg[ C+ \ln(\frac{H}{\mu}) \bigg]~,
\end{eqnarray}
where we have used $aH=q$ at the moment of Hubble-crossing. $\Lambda_{UV}$ is a physical cutoff at UV regime, and $C$ is some integral constant which is of $O(1)$.

From the result \eqref{eq:massive-loop-final}, one can find that the divergence of leading order is the UV divergence, and there is a mixture between the UV and IR divergence due to the logarithmic term at the next-to-leading order. Moreover, this loop correction is proportional to the factor $\frac{m}{H}$ and thus becomes negligible when the fermion field is much lighter than the inflaton. Eventually, since the final integral is proportional to $q^{-3}$, it implies that the scale invariance of the primordial tensor fluctuations would not be changed under the fermion loop correction.

\section{Conclusion}

To conclude, we in the current paper have studied the loop corrections of a fermion field to the primordial power spectrum of tensor fluctuations in the frame of inflationary cosmology. Our calculation is based on the {\it in-in} formalism and the spinor field is assumed to be a Dirac fermion. In this setup, we have separately considered the loop contribution of a massless fermion and that of a massive fermion. For the case of a massless fermion, we obtain a logarithmic correction in the correlation function of primordial tensor fluctuations which involves a regularization parameter in IR regime. This result is in agreement with previous works\cite{Weinberg:2005vy, Chaicherdsakul:2006ui}. Notice that, our result shows that this IR issue is irrelevant to the scale factor rescaling. Thus the IR regularization parameter is a physical one, which is consistent with the analysis in Ref. \cite{Senatore:2009cf}.

Further, we have extended our calculation to the case of a massive fermion field. We find that the inclusion of a fermion's mass can be helpful to relax the IR issue since the logarithmic function obtained in the massless fermion loop is replaced by a $\cosh$ function which is exponentially suppressed at large mass limit. Moreover, at the small mass limit, the fermion correction is also suppressed since the loop integral is proportional to $m^2/H^2$. However, we should be aware of that this result is based on the assumption that in the loop integral most of the perturbation modes are in the super-Hubble scale. It is interesting to study in detail the contribution of the sub-Hubble modes in the loop integral and see how it would affect the IR issue in the case of a massive fermion field. We would like to leave it as a follow-up project.

Moreover, due to the fermion's mass term, one can obtain a fermion loop to the correlation function of primordial tensor fluctuations with single vertex. In usual quantum field theory, this term can also appear but only contribute to the UV renormalization. However, in the context of inflationary cosmology, we find the leading term of the one-loop correction is proportional to $\Lambda_{UV}^3$ while the next-to-leading term is a mixture of the UV and IR divergences. It is necessary to continue the study in this issue and see if there is any implication on the relation between inflation models and quantum gravity. We notice that the authors of Ref. \cite{Chen:2012ye} also acquired a similar result on primordial curvature perturbation in the presence of loop corrections of a second entropy field.

As an ending remark, we would like to point out all the above loop corrections do not change the scale dependence of the correlation function of primordial tensor modes. Therefore, we expect a nearly scale-invariant power spectrum of primordial tensor fluctuations.

\textbf{Acknowledgments}

We would like to thank Wei Xue for helpful discussions and valuable comments on the manuscript. The work of FKX and PYS is supported in part by NSFC under Grant No:10775180, 11075205, in part by the Scientific Research Fund of GUCAS (NO:055101BM03), in part by National Basic Research Program of China, No:2010CB832804. The work of CYF is supported in part by the Cosmology Initiative in Arizona State University. One of the author (CYF) is grateful to the other two for hospitality during his visit in GUCAS. And he also thanks Xinmin Zhang in the Institute of High Energy Physics at CAS for hospitality while this work was finalized.

\end{document}